\documentclass[aps,prl,reprint,superscriptaddress,nopacs,floatfix, showkeys]{revtex4-1}

\usepackage{graphicx, color}
\usepackage{amssymb, amsmath}
\usepackage{times}
\usepackage[normalem]{ulem}
\usepackage{bbm}
\usepackage{todonotes}
\usepackage{mathtools}
\usepackage{enumitem} 
\usepackage{physics}
\usepackage{braket}
\usepackage{soul,xcolor}
\usepackage{booktabs}  
\usepackage{tabularx}

\makeatletter
\newcommand{\thickhline}{%
    \noalign {\ifnum 0=`}\fi \hrule height 1pt
    \futurelet \reserved@a \@xhline
}

\newcolumntype{"}{@{\hskip\tabcolsep\vrule width 1pt\hskip\tabcolsep}}
\renewcommand{\t}{\text} 
 
\newcommand{\p}[1]{\left(#1\right)}

\makeatother

\begin{document}

\title{
Emergence of multi-body interactions in few-atom sites of a fermionic lattice clock
}

\date{\today}
\author{A. Goban}
\email{These authors contributed equally to this work.}
\affiliation{JILA, National Institute of Standards and Technology and University of Colorado, 440 UCB, Boulder, Colorado 80309, USA}
\affiliation{Department of Physics, University of Colorado, 390 UCB, Boulder, Colorado 80309, USA}

\author{R. B. Hutson}
\email{These authors contributed equally to this work.}
\affiliation{JILA, National Institute of Standards and Technology and University of Colorado, 440 UCB, Boulder, Colorado 80309, USA}
\affiliation{Department of Physics, University of Colorado, 390 UCB, Boulder, Colorado 80309, USA}

\author{G. E. Marti}
\affiliation{JILA, National Institute of Standards and Technology and University of Colorado, 440 UCB, Boulder, Colorado 80309, USA}
\affiliation{Department of Physics, University of Colorado, 390 UCB, Boulder, Colorado 80309, USA}

\author{S. L. Campbell}
\email{Present address: Molecular Biophysics and Integrated Bioimaging, Lawrence Berkeley National Laboratory, Berkeley, CA 94720, USA; Department of Physics, University of California, Berkeley, CA 94720, USA}
\affiliation{JILA, National Institute of Standards and Technology and University of Colorado, 440 UCB, Boulder, Colorado 80309, USA}
\affiliation{Department of Physics, University of Colorado, 390 UCB, Boulder, Colorado 80309, USA}

\author{\\M. A. Perlin}
\affiliation{JILA, National Institute of Standards and Technology and University of Colorado, 440 UCB, Boulder, Colorado 80309, USA}
\affiliation{Department of Physics, University of Colorado, 390 UCB, Boulder, Colorado 80309, USA}
\author{P. S. Julienne}
\affiliation{Joint Quantum Institute, NIST and the University of Maryland, Gaithersburg, Maryland 20899-8423, USA}
\author{J. P. D'Incao}
\affiliation{JILA, National Institute of Standards and Technology and University of Colorado, 440 UCB, Boulder, Colorado 80309, USA}
\affiliation{Department of Physics, University of Colorado, 390 UCB, Boulder, Colorado 80309, USA}
\author{A. M. Rey}
\affiliation{JILA, National Institute of Standards and Technology and University of Colorado, 440 UCB, Boulder, Colorado 80309, USA}
\affiliation{Department of Physics, University of Colorado, 390 UCB, Boulder, Colorado 80309, USA}
\author{J. Ye}
\affiliation{JILA, National Institute of Standards and Technology and University of Colorado, 440 UCB, Boulder, Colorado 80309, USA}
\affiliation{Department of Physics, University of Colorado, 390 UCB, Boulder, Colorado 80309, USA}

\begin{abstract}
Alkaline-earth (AE) atoms have metastable clock states with minute-long optical lifetimes, high-spin nuclei, and SU($N$)-symmetric interactions that uniquely position them for advancing atomic clocks \cite{Bloom:2013uoa,Nicholson:2015kq, Ushijima:2015gaa, Schioppo:2016fj}, quantum information processing \cite{Daley:2011cv}, and quantum simulation \cite{Cazalilla:2014kq}. 
The interplay of precision measurement and quantum many-body physics is beginning to foster an exciting scientific frontier with many opportunities \cite{Martin:2013kl,Campbell:2017ic}. 
Few particle systems provide a window to view the emergence of complex many-body phenomena arising from pairwise interactions \cite{Wenz:2013er}.
Here, we create arrays of isolated few-body systems in a fermionic ${}^{87}$Sr three-dimensional (3D) optical lattice clock and use high resolution clock spectroscopy to directly observe the onset of both elastic and inelastic multi-body interactions.
These interactions cannot be broken down into sums over the underlying pairwise interactions. 
We measure particle-number-dependent frequency shifts of the clock transition for atom numbers $n$ ranging from 1 to 5, and observe nonlinear interaction shifts, which are characteristic of  SU($N$)-symmetric elastic multi-body effects.
To study inelastic multi-body effects, we use these frequency shifts to isolate $n$-occupied sites and measure the corresponding lifetimes.
This allows us to access the short-range few-body physics free from systematic effects encountered in a bulk gas. 
These measurements, combined with theory, elucidate an emergence of multi-body effects in few-body systems of sites populated with ground-state atoms and those with single electronic excitations.
By connecting these few-body systems through tunneling,
the favorable energy and timescales of the interactions will allow our system to be utilized 
for studies of high-spin quantum magnetism \cite{Zhang:2014elb} and the Kondo effect \cite{FossFeig:2010haa,Isaev:2016fe, Cazalilla:2014kq, Isaev:2015cd, Nakagawa:2015ev, Riegger:2017wq}.
\end{abstract}

\maketitle

Fermionic AE and AE-like atoms have ground ${}^1\mathrm{S}_0$ and long-lived metastable  ${}^3\mathrm{P}_0$ ($\sim$160 s lifetime for ${}^{87}\mathrm{Sr}$) ``clock'' states, 
which provide two (electronic) orbital degrees of freedom that are largely decoupled from the nuclear spin, $I$.
This gives rise to orbital SU($N=2I+1$)-symmetric two-body interactions where the $s$-wave and $p$-wave scattering parameters are independent of the nuclear spin state \cite{Gorshkov:2010hw, Cazalilla:2009ir, Cazalilla:2014kq}.
This degeneracy can be quite large ($I = 9/2$ for ${}^{87}\mathrm{Sr}$) and thus enables studies of quantum states of matter with no direct analogues in nature, such as the SU($N$) Mott insulator~\cite{Taie:2012jy, Hofrichter:2016iq, Cazalilla:2014kq}.
Two-orbital, SU($N$)-symmetric interactions were first directly observed through clock spectroscopy~\cite{Zhang:2014elb, Scazza:2014jw, Cappellini:2014fv}, and have since enabled new opportunities for studying strongly interacting Fermi gases~\cite{Kato:2013jq, Hofer:2015ir, Pagano:2015il, Ozawa:2018as} and the Kondo lattice model~\cite{Riegger:2017wq}.

While particles microscopically interact pairwise, multi-body interactions can emerge in a low-energy effective field theory where fluctuations beyond some length, or momentum scale, are ``integrated out.''
Examples of this include three-nucleon forces~\cite{Hammer:2013kn} and some fractional quantum Hall states~\cite{Fradkin:1997ge}.
Such multi-body interactions have been predicted to arise in a variety of optical lattice experiments ~\cite{Buchler:2007dk, Johnson:2009bj}, and have been observed in bosonic systems~\cite{Will:2010dm, Mark:2011kl, Franchi:2017ue}.
While a single impurity interacting with a few identical fermions has been studied \cite{Wenz:2013er}, multi-body interactions in high-spin fermions have thus far remained unexplored.

In ultracold gases, the effects of multi-body interactions have also been extensively explored in the context of three-body recombination processes~\cite{Ferlaino:2011fw, wang2013Adv, anrev:am}.
These include studies of exotic Efimov states and other forms of universality associated with long range interactions \cite{Braaten:2006dl,Naidon:2017km, Wang:2014hx,Greene:2017fz, DIncao:2017tg}.
However, comparison to theory has been often difficult due to the bulk gas nature of these experiments.
Improved control and understanding of the atoms' external degrees of freedom is crucial for testing theoretical models of ultracold collisions~\cite{Wang:2012bc, Wolf:2017fc}.

Here we study the emergence of multi-body interactions by combining isolated few-body systems in an optical lattice with high resolution clock spectroscopy.
In this experiment, the preparation of the ultracold gas proceeds similarly to Refs. \cite{Campbell:2017ic, Marti:2017vq}.
In summary, we prepare a 10-spin-component Fermi degenerate gas, with atoms equally distributed amongst all nuclear spin states. Typically we produce $10^3-10^4$ atoms per nuclear spin state at a temperature $T=10-20~{\rm nK}=0.1~T_F$, where $T_F$ is the Fermi temperature. 
The gas is loaded into a nearly isotropic 3D optical lattice where the geometric mean of the trap depths for the three lattice beams, $\mathcal{U}$, varies from $30$ to $80~E_{\rm rec}$, where $E_{\rm rec} = h \times 3.5\,\mathrm{kHz}$ is the lattice photon recoil energy.
At these trap depths, there is negligible tunneling between neighboring sites over the timescale of the experiment.

\begin{figure}[t!]
\centering
\includegraphics[width=1\columnwidth]{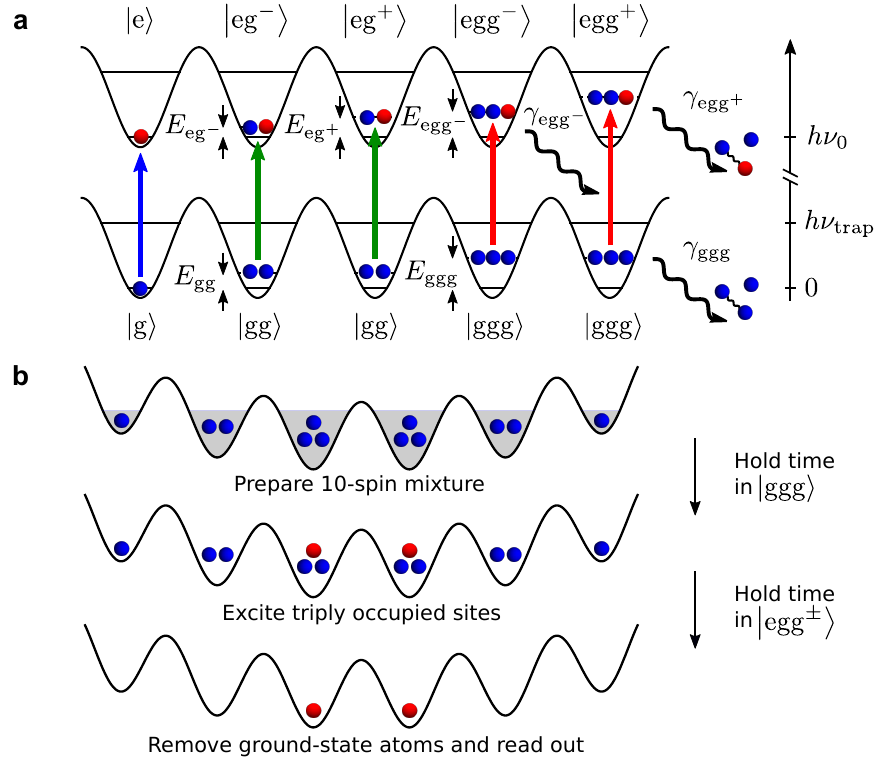}
\caption{
    \textbf{Two-orbital interactions in a 3D lattice and experimental sequence.} 
    \textbf{a,} One to three atoms occupy the lowest motional state of a lattice site with corresponding on-site energies $E_X$. 
    We use a state-independent lattice operating at the magic wavelength where the polarizabilities of the electronic ground and excited states are identical.
    In a deep 3D lattice, each site can be regarded as an isolated few-body system.
    A clock photon resonantly couples the ground state $\ket{\mathrm{g}\cdots}$ to the single-excitation manifold $\ket{\mathrm{eg}\cdots^{\pm}}$, leading to a spectroscopic shift from the bare resonance frequency $\nu_0\approx$ 429 THz. 
    Multi-body interactions manifest themselves in sites with three or more atoms both in the observed clock shifts, and in their decay into a diatomic molecule plus a free atom at a rate $\gamma_X$.
    \textbf{b,} Experimental sequence for imaging triply occupied sites.
    A 10 nuclear-spin mixture is loaded into a 3D optical lattice. 
    A clock pulse resonantly drives triply-occupied sites $\ket{\mathrm{ggg}}$ to an excited state $\ket{\mathrm{egg}^{\pm}}$. 
    After all atoms in the ground state are removed, the atoms remaining in the excited state are read out with absorption imaging. 
    Three-body decay rates are measured by adding a hold time before (for $\gamma_{\mathrm{ggg}}$) or after applying the clock pulse (for $\gamma_{\mathrm{egg}^{\pm}}$).
}
\label{fig:fig1}%
\end{figure}

As depicted in Fig.\ref{fig:fig1}a, for atoms in doubly occupied sites, a $\pi$-polarized clock photon resonantly couples the ground state $|\mathrm{gg}\rangle$ to the orbital-symmetric (anti-symmetric) excited state $\ket{\mathrm{eg}^{+(-)}}$ upon matching the detuning, $(E_{\mathrm{eg}^{+(-)}} - E_\mathrm{gg} ) / h$, at zero magnetic field.
Here, $\mathrm{g}(\mathrm{e})$ represents the ${}^{1}\mathrm{S}_0$ $({}^{3}\mathrm{P}_0)$ clock state,  $E_X$ is the on-site interaction energy for $X\in\{\mathrm{gg}, \mathrm{eg}^{\pm}\}$, and $h$ is the Planck constant.
Similarly, for sites with $n\geq 3$, the ground state $\ket{\mathrm{g}\cdots}$ can be driven to the state $\ket{\mathrm{eg}\cdots^{+}}$, which is an orbitally symmetric state, or the state $\ket{\mathrm{eg}\cdots^-}$, for which the orbital and nuclear spin degrees of freedom are not separable.
The $\pi$-polarized clock light preserves the initial nuclear-spin state distribution.

\begin{figure}[t]
\centering
\includegraphics[width=1\columnwidth]{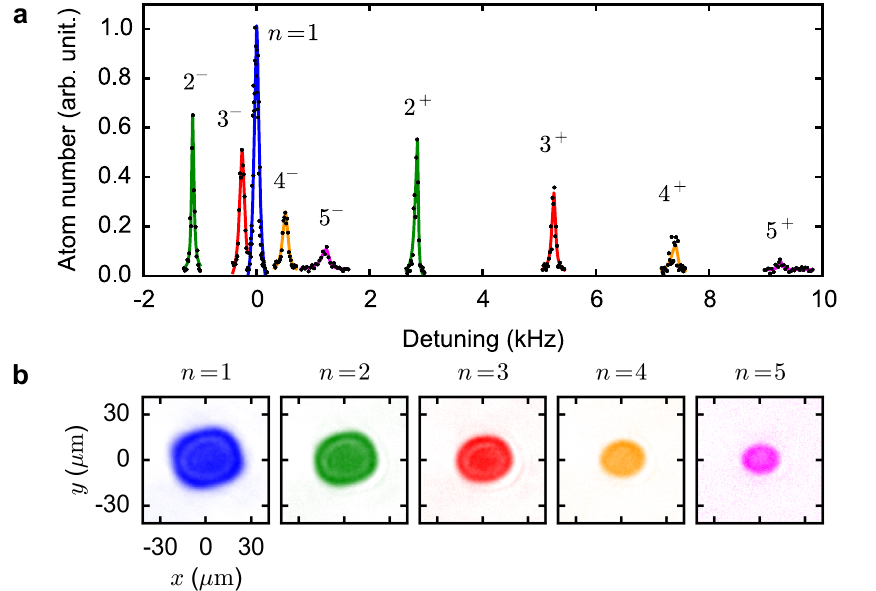}
\caption{
    \textbf{Clock spectroscopy of a 10-component Fermi gas in a 3D lattice.} 
    \textbf{a,} Overlayed clock spectra for occupation numbers $n=1, \ldots, 5$ at a mean trap depth $\mathcal{U} = 54\; E_\mathrm{rec}$ ($\nu_{\mathrm{trap}}=51$ kHz).
    The labels $n^{\pm}$ denote excitation of $\ket{\mathrm{eg}\cdots^\pm}$ for $n$-occupied sites.
    For large occupations, the line shapes become asymmetric due to the inhomogeneity of the trap depth.
    The solid lines are fits used to determine the resonance frequencies (see Methods).
    The detunings are given relative to the resonance of the clock transition for singly occupied sites (blue).
    Each data point is the result of a single experimental cycle.
    \textbf{b,} Column densities of different occupation numbers for a sample of $2 \times 10^5$ atoms. 
    The absorption images for different occupation numbers are taken according to the procedure in Fig. \ref{fig:fig1}\textbf{b}, by first exciting on the symmetric resonances. 
    Each image is averaged over $20$ experimental cycles.}
\label{fig:fig2}%
\end{figure}

We spatially resolve the spectroscopic signal using absorption imaging~\cite{Marti:2017vq} and the readout scheme presented in Fig.~\ref{fig:fig1}b.
We measure both the differential interaction energies and the spatial distributions of each occupation number \cite{Campbell:2006im, Franchi:2017ue, Kato:2016bm, Bouganne:2017wq}.
Fig.~\ref{fig:fig2}a shows sample spectra of a 10-spin-component Fermi gas using 20 ms clock pulses from a 26 mHz-linewidth ultrastable laser.
For each occupation number $n$, there is a pair of single-excitation resonances, labeled $n^\pm$, corresponding to the two sets of final states, $|\mathrm{eg}\cdots^\pm\rangle$.
SU($N$) symmetry and fermionic anti-symmetrization dictate that only two eigenenergies appear for each $n$-atom sample (see Methods).

Figure~\ref{fig:fig2}b shows the column density of different occupation numbers for a sample of $2\times 10^5$ atoms.
The shells of decreasing size with increasing occupation number are a result of balancing the external confinement generated by Gaussian lattice beams with the on-site interaction energies. 
As observed for small $n$, larger clouds of atoms extend over areas where the trapping frequencies are relatively lower, resulting in smaller on-site interaction energies.
To eliminate a possible systematic shift from the changing cloud size, we adjust the final evaporation point to maximize a central density of the desired occupation number and measure the spectroscopic response in only the central 4 $\mu$m $\times$ 4 $\mu$m $\times$ 2 $\mu$m region of the trap.
The vertical plane is selected by loading the lattice from a trap that is tightly confining against gravity, loading only a 2 $\mu$m-thick vertical region.
Spatial selection in the horizontal plane is performed by spatially filtering the images, measuring the response from only the central region of the lattice.
The trap depth in the central region of the lattice is calibrated via motional sideband spectroscopy of a $n=1$ sample with the same spatial selection.
We note that the current images show in-plane density distributions, integrated along the imaging axis.

To investigate multi-body interactions in multiply occupied sites, we first consider the case of two interacting fermionic atoms, each with two internal degrees of freedom: an electronic orbital, $x \in \{\mathrm{g}, \mathrm{e}\}$, and a nuclear spin sublevel, $m \in \{-I, -I + 1, \ldots, I\}$. 
The interactions depend only on the electronic degree of freedom, so all $s$-wave scattering processes are parameterized by four scattering lengths, $a_X$, with $X \in \{\mathrm{gg}, \mathrm{eg}^+, \mathrm{eg}^-, \mathrm{ee}\}$, resulting in SU($N$) symmetric properties of the system.
Here, $+$($-$) denotes a symmetric (anti-symmetric) superposition of the electronic orbitals. 
In our experiments, atoms are trapped in the motional ground states of deep lattice sites with a single-particle Wannier function, $\phi_0(\mathbf{r})$, localized to a characteristic length scale, $l_0$.
Since all atoms are in the ground motional states, the Pauli exclusion principle requires that atoms with the same orbital state, $x$, have different nuclear spins, $m$. Here, we consider the case where each atom is in a different spin state.
In the limit of weak interactions ($l_0 \gg |a_X|$), the pairwise interaction energy can be expressed as $U_X^{(2)} = \frac{4 \pi \hbar^2}{m_a} a_X \int \mathrm{d}^3 \mathbf{r} \left|\phi_0(\mathbf{r})\right|^4$, where $m_a$ is the atomic mass, and $\hbar=h/2\pi$.
In this regime, the on-site many-body Hamiltonian is described by, 
\begin{equation}
\begin{aligned}
    H =& \sum_{m\neq m'}\Big[\frac{U^{(2)}_{\mathrm{gg}}}{2}n_{\mathrm{g}, m} n_{\mathrm{g}, m'}  + \frac{U^{(2)}_{\mathrm{ee}}}{2}n_{\mathrm{e}, m}n_{\mathrm{e}, m'}\\ 
    +& V^{(2)}n_{\mathrm{e}, m} n_{\mathrm{g}, m'} + V^{(2)}_\mathrm{ex} c^\dag_{\mathrm{e}, m} c^\dag_{\mathrm{g}, m'}c_{\mathrm{e}, m'} c_{\mathrm{g}, m}\Big],
\label{eqn:two-body-hamiltonian}
\end{aligned}
\end{equation}
where $c^\dagger_{x, m}$ ($c_{x,m}$) creates (destroys) an atom in orbital $x\in\{\mathrm{e}, \mathrm{g}\}$ with spin $m$, and $n_{x, m} = c^\dagger_{x,m}c_{x,m}$. The direct and exchange interaction energies are $V^{(2)} = (U^{(2)}_{\mathrm{eg}^+} + U^{(2)}_{\mathrm{eg}^-}) / 2$, and $V^{(2)}_\mathrm{ex} = (U^{(2)}_{\mathrm{eg}^+} - U^{(2)}_{\mathrm{eg}^-}) / 2$, respectively.
The Hamiltonian in equation (\ref{eqn:two-body-hamiltonian}) has a ground state $\ket{\mathrm{g}\cdots}$ with a corresponding eigenenergy $E^{(2)}_{\mathrm{g}\cdots}$ and two distinct eigenenergies $E^{(2)}_{\mathrm{eg}\cdots^{\pm}}$ with a single excitation, one for an orbital-symmetric excited state $\ket{\mathrm{eg}\cdots^+}$ and the other one for $(n-1)$-fold degenerate excited states $\ket{\mathrm{eg}\cdots^-}$ (see Methods). 
We note that this orbital-symmetric state is an $n$-body entangled $W$ state \cite{Zang:2015kb}.

For tighter confinement and stronger on-site interactions, i.e. if $a_X/l_0$ is not negligible, corrections to equation (\ref{eqn:two-body-hamiltonian}) become increasingly important.
The increased interaction energy facilitates off-resonant transitions to higher motional states.
Equivalently, the spatial wave function $\phi_0(\mathbf{r})$ becomes dependent on the number of atoms per site and their configuration.
This effect can be captured by a lowest-band effective Hamiltonian where the higher motional states are integrated out, with two consequences with (i) The two-body interaction energies are characterized by an \textit{in-trap} scattering length, rescaled from the free-space one. (ii) The total interaction energy for $n\geq3$ atoms cannot be broken down into a sum over pairs of atoms~\cite{Johnson:2012cl}, leading to effective multi-body interactions.
Considering at most one atom in the excited state, equation (\ref{eqn:two-body-hamiltonian}) must be modified to include multi-body corrections,
\begin{equation}
\begin{aligned}
    H' = & \sum_{m\neq m'\neq m''} \Bigg[\frac{U_{\mathrm{ggg}}^{(3)}}{6} n_{\mathrm{g}, m}n_{\mathrm{g}, m'}n_{\mathrm{g}, m''} \\
    \quad & + \frac{V^{(3)}}{2} n_{\mathrm{e}, m}n_{\mathrm{g}, m'}n_{\mathrm{g}, m''} \\
    \quad & +  \frac{V_\mathrm{ex.}^{(3)}}{2} c^\dag_{\mathrm{e}, m'} c^\dag_{\mathrm{g}, m}c_{\mathrm{e}, m} c_{\mathrm{g}, m'} n_{\mathrm{g}, m''}\Bigg] + \mathcal{O}(n^4),
\label{eqn:multi-body-hamiltonian}
\end{aligned}
\end{equation}
where $U^{(3)}_\mathrm{ggg}$, $V^{(3)}$ and $V^{(3)}_\mathrm{ex}$ are effective three-body ground-state, direct, and exchange interaction energies.
Due to the SU($N$) symmetry, $H'$ has the same eigenstates as $H$, but with modified $n$-body eigenenergies (See Methods and Ref. \cite{Perlin:2017va}).
These multi-body interactions can be probed by spectroscopically addressing lattice sites with different occupation numbers.

\begin{figure}[t!]
\centering
\includegraphics[width=1\columnwidth]{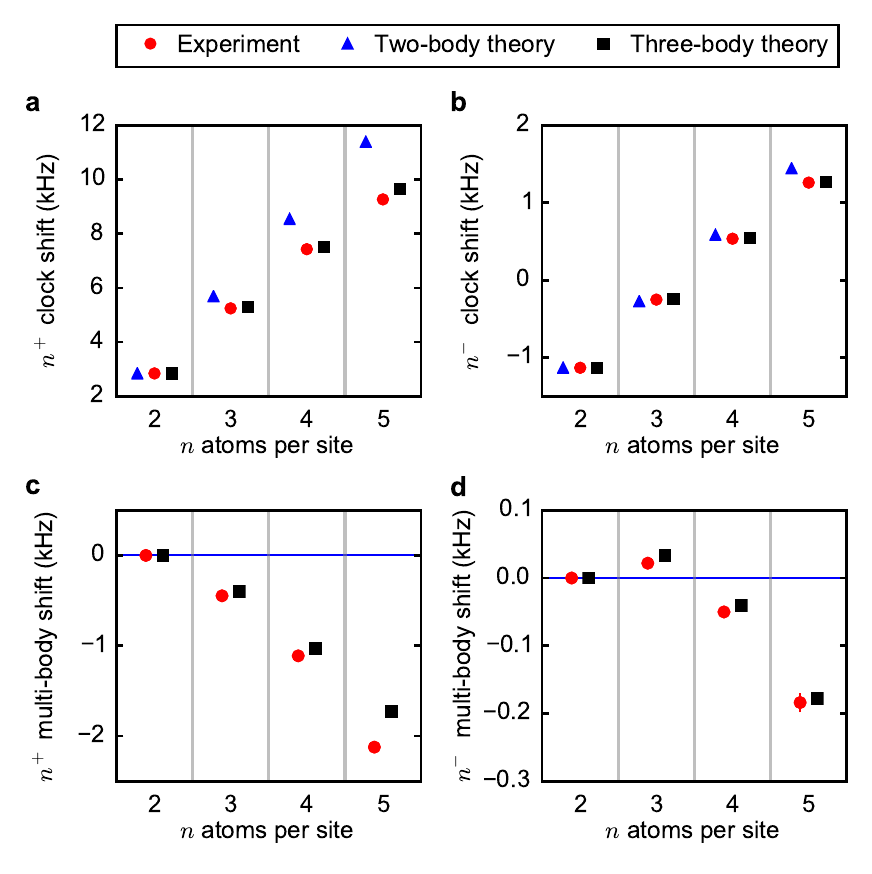}
\caption{
    \textbf{Effective multi-body clock shifts.} 
   \textbf{a,} Clock shifts of $\ket{\mathrm{eg}\cdots^+}$ from $n=2,3,4, \mathrm{and}~5$, at a mean trap depth of $\mathcal{U}=54E_{\rm rec}$ ($\nu_{\rm trap}=51$ kHz).
    Effective multi-body interactions are observed in the experimental data (red circles) as a deviation from the two-body prediction (blue triangles).
    The calculated shifts from an effective Hamiltonian including three-body interactions (see text and Methods) are shown in black squares.
    The points at a given occupation number are horizontally offset for clarity. 
    The uncertainties of the experimental data are smaller than the size of the data points.
   \textbf{b,} Clock shifts of $\ket{\mathrm{eg}\cdots^-}$ taken at the same conditions as in \textbf{a}.
    The two-body theory shows smaller deviations from the measured shifts at $n=3$ and $4$ due to a near cancellation of the three-body shifts between $|\mathrm{g}\cdots\rangle$ and $
    \ket{\mathrm{eg}\cdots^-}$.
   \textbf{c, d,} Multi-body interaction shifts where the two-body contributions are subtracted from the points in \textbf{a} and \textbf{b}.
}
\label{fig:fig3}
\end{figure}

To extract the multi-body effects from equation (\ref{eqn:multi-body-hamiltonian}), we first measure the frequency shifts $(E_{\mathrm{eg}^\pm} - E_\mathrm{gg})/h$ for various mean trap depths.
By incorporating the corrections of lattice confinement with a previous measurement of the ground state scattering length, $a_{\mathrm{gg}} = 96.2(0.1)a_0$ where $a_0$ is the Bohr radius \cite{MartinezdeEscobar:2008eu, Stein:2010dg}, we extract the free-space scattering lengths $a_{\mathrm{eg}^{\pm}}$, shown in Table \ref{tab:experiments} (see Methods).

Multi-body interactions occur only in sites with three or more atoms and cause frequency shifts that are nonlinear in occupation number $n$.
The measured clock shifts of the $|\mathrm{eg} \cdots^+ \rangle$ ($|\mathrm{eg} \cdots^- \rangle$) branch are shown as red points in Fig.~\ref{fig:fig3}a(b).
They show deviations from the values expected from equation (\ref{eqn:two-body-hamiltonian}) (blue triangles) that are proportional to the occupation number $n$, and are consistent with the three-body corrections in equation (\ref{eqn:multi-body-hamiltonian}) (black squares).
These higher-order contributions, shown in Fig. \ref{fig:fig3}c and \ref{fig:fig3}d can be intuitively interpreted as broadening of the wave function, lowering the magnitude of overall interaction energy.
From a variational calculation, we find that the wave function of $n=5$ atoms is broadened by $\sim 8$\% relative to a non-interacting one.

Multi-body effects also appear in these few-body systems as three-body recombination loss. 
These losses occur when three atoms recombine to form a deeply-bound diatomic molecule and a free-atom, both carrying enough energy to eject them from the trap \cite{Wolf:2017fc}.
We selectively determine the lifetime of a given $n$-atom $|\mathrm{g}\cdots\rangle$ state by holding atoms in a deep lattice for a variable time, then resonantly driving $|\mathrm{g}\cdots\rangle \rightarrow |\mathrm{eg}\cdots^\pm\rangle$ to spectroscopically address only the $n$-atom sites, and finally measuring the $e$-atom population after removing the $g$ atoms, as illustrated in Fig. \ref{fig:fig1}. 
Similarly, the loss rate of the $|\mathrm{eg}\cdots^\pm\rangle$ state is determined by first driving $|\mathrm{g}\cdots\rangle \rightarrow |\mathrm{eg}\cdots^\pm\rangle$, and then holding for a variable time before the $g$-atom removal.
In both cases, we fit an exponential decay to the measured excited state atom number at the end of the experimental sequence.
This analysis is dramatically simpler than that for bulk-gas experiments where decay curves must be fit with multiple rate constants corresponding to one-, two-, and three-body losses \cite{Burt:1997zz, Soding:1999jq}.

\begin{figure}[t!]
\centering
\includegraphics[width=1\columnwidth]{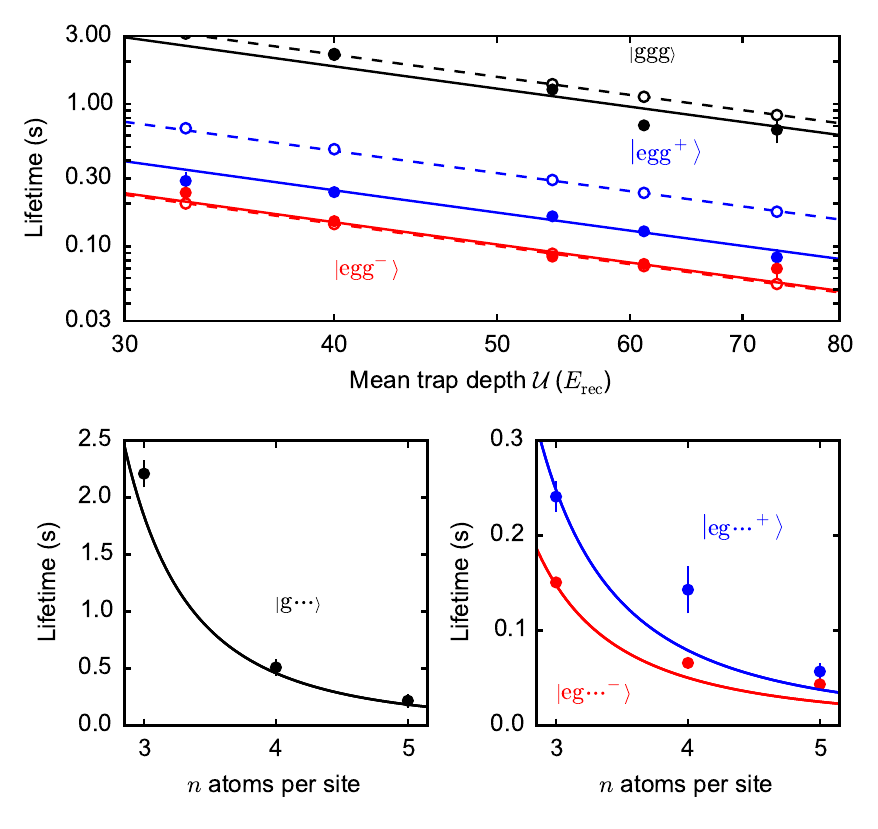}
\caption{
    \textbf{Three-body loss rate and occupation-number dependent lifetime.} \textbf{a,} Mean trap depth dependence of $n=3$ lifetimes in $\left|\mathrm{ggg}\right>$ (black), $\left|\mathrm{egg}^{+}\right>$ (blue) and $\left|\mathrm{egg}^{-}\right>$ (red).
    The measured lifetimes (closed circles) are close to the calculated ones (open circles) from a universal van der Waals model (see text and Methods).
    From the fits shown in solid (dashed) lines, we extract three-body loss coefficients $\beta_X$ for the measured (calculated) lifetimes, summarized in Table \ref{tab:experiments}.
    The lifetime of $\ket{\mathrm{ggg}}$ is ten times longer than that of $\ket{\mathrm{egg}^{\pm}}$, since the number of molecular states increases due to one distinguishable particle in the $e$ state.
    \textbf{b, c,} Occupation-number dependence of the $\ket{\mathrm{g}\cdots}$ and $\ket{\mathrm{eg}\cdots^{\pm}}$ lifetimes at $\mathcal{U}=40\,E_{\rm rec}$.
    The solid lines are calculated lifetime assuming pure three-body losses where measured $\beta_{\mathrm{ggg}}$ and $\beta_{\mathrm{egg}^{\pm}}$ as the input parameters (see text and Methods)
}
\label{fig:fig4}
\end{figure}

To disentangle these multi-body effects from inelastic two-body collisions, we first measure the ground and excited state lifetimes of the one- and two-atom sites.
While we observe a vacuum limited, ${\sim}100\ \mathrm{s}$, $ 1 / e$ lifetime for the $|\mathrm{g}\rangle$ and $|\mathrm{gg}\rangle$ states, off-resonant Raman scattering from the optical lattice light causes a decay of the single atom excited state, $\ket{\mathrm{e}} \rightarrow \ket{\mathrm{g}}$, with a $9.6(0.4)\ \mathrm{s}$ time constant at a mean trap depth of $\mathcal{U} = 73\ E_\mathrm{rec}$~\cite{Dorscher:2018uj, Hutson:2018}.
At this same trap depth, we find the lifetimes, $\tau_{\mathrm{eg}^\pm}$, of $|\mathrm{eg}^+\rangle$ and $|\mathrm{eg}^-\rangle$ to be $5.1(0.7)\ \mathrm{s}$ and $6.1(0.7)\ \mathrm{s}$, respectively.
Such two-body lifetimes can be related to a two-body loss coefficient via the expression $\tau_{\mathrm{eg}^{\pm}}^{-1}= \beta_{\mathrm{eg}^{\pm}}\int{\rm d}^3\mathbf{r} |\phi_0({\bf r})|^4$ \cite{GarciaRipoll:2009fa}.
However, since the two-body lifetimes are only slightly shorter than that of a single excited atom, we can only determine $\beta_{\mathrm{eg}^+} \leq 2.5(0.3) \times 10^{-16}\,\mathrm{cm}^3/\mathrm{s}$ and $\beta_{\mathrm{eg}^-} \leq 2.1(0.2) \times 10^{-16}\,\mathrm{cm}^3/\mathrm{s}$ as upper limits.

The measured lifetimes of the three-atom states, $\tau_X$ for $X \in \{\mathrm{ggg}, \mathrm{egg}^+, \mathrm{egg}^-\}$, at various mean trap depths are shown in Fig. \ref{fig:fig4}a.
These multi-body decays all occur on timescales significantly shorter than those of one- and two-body losses.
Furthermore, the excited states are observed to decay faster than the ground state by approximately an order of magnitude.
We attribute this to the increased number of molecular decay channels after replacing a $g$ atom with a distinguishable $e$ atom.
Table \ref{tab:experiments} shows the density-independent three-body loss coefficient $\beta_X$ extracted from these measurements via the expression $\tau_X^{-1} = \beta_X \int \mathrm{d}^3 \mathbf{r} |\phi_0(\mathbf{r})|^6$~\cite{Jack:2003ey}.

Next, we compare the measured three-body lifetimes to a model in which atoms interact by pairwise additive long-range van der Waals potentials joined at shorter range to a pseudopotential that is adjusted to yield in each case the measured two-body scattering lengths given in Table \ref{tab:experiments}~\cite{Zhang:2014elb,Wang:2012bc,Wang:2014hx,Wolf:2017fc}.
Numerically solving the three-body Schr{\" o}dinger equation yields the frequency shifts and the decay lifetimes for three atoms confined in a harmonic trap (see Methods) \cite{Blume:2012ct}.
We increase the number of bound states in each pairwise potential  until all results converge at the $<10\%$ level.
As shown in Fig. \ref{fig:fig4}a, the calculated lifetimes (open circles) are remarkably close to the measured lifetimes (closed circles), given the simplicity of our universal van der Waals model with no fit parameters. 
We note that while our results for $\ket{\mathrm{ggg}}$ and $\ket{\mathrm{egg}^-}$ agree with the observed lifetimes to within $<15\%$, the results for $\ket{\mathrm{egg}^+}$ overestimate the lifetimes by about $50\%$, most likely due to the fact that for this state our model does not allow for decay into all possible diatomic molecular states (see Methods).
As a sanity check, the frequency shifts produced by this model agree with the measurements shown in Fig. \ref{fig:fig3}a, b to within 10\%, despite assuming a harmonic trap potential.

\begin{table}[t]
\def\arraystretch{1.25}%
\begin{tabularx}{240pt}{@{}ccc@{}}
\thickhline
~ Channel & ~~$s$-wave scattering lengths & ~~Two-body loss coefficients  \\
$X$  & $a_X(a_0)$ & $\beta_X \left(10^{-16}{\rm cm}^3/{\rm s}\right)$  \\ 
 \hline
 $\mathrm{gg}$ &  $96.2(0.1)$  \\
$\mathrm{eg}^{-}$ &  $\quad 69.1(0.2)_{\rm stat}(0.9)_{\rm sys}$ & $\quad\leq 2.1(0.2)$ \\
$\mathrm{eg}^{+}$ & $\quad 160.0 (0.5)_{\rm stat}(2.3)_{\rm sys}$& $\quad\leq 2.5(0.3)$\\
\thickhline 
~ Channel & \multicolumn{2}{c}{~~Three-body loss coefficients: $\beta_X \left(10^{-30}{\rm cm}^6/{\rm s}\right)$}\\
$X$ & Measured  & Calculated  \\ 
 \hline 
$\mathrm{ggg}$ & 2.0 (0.2) & 1.7 \\
$\mathrm{egg}^{-}$ &   25 (1) & 26 \\
$\mathrm{egg}^+$  &  15 (1) & 8.0 \\
\thickhline
\end{tabularx}
\caption{\textbf{$s$-wave scattering lengths and three-body loss coefficients.} 
The scattering length of ground states $a_\mathrm{gg}=96.2(0.1)a_0$ is determined from photoassociation spectroscopy \cite{MartinezdeEscobar:2008eu, Stein:2010dg}, while all the other values are extracted in this work. 
The measured elastic $s$-wave scattering lengths are consistent with previous reported values in Ref. \cite{Zhang:2014elb}, with a 10-fold improvement for the uncertainty of $a_{\mathrm{eg}^-}$. The two-body loss coefficients are upper bounds, limited by $\ket{\mathrm{e}}$ state lifetime. 
The measured three-body loss coefficients are in good agreement with the calculated ones based on a universal van der Waals model.
}
\label{tab:experiments}
\end{table}

We extract three-body loss coefficients $\beta_X$ from the calculated lifetimes by the same procedure as for the experimental results shown in Table \ref{tab:experiments}. 
The good agreement of the universal van der Waals model with our ${}^{87}\mathrm{Sr}$ lattice experiment is in sharp contrast to the disagreement, by a factor of 2 to 4, for bulk-gas ${}^{87}\mathrm{Rb}$ experiments \footnote{The universal van  model gives a three-body loss rate coefficient $L_3=1.0\times10^{-29}$ cm$^6$/s \cite{Wolf:2017fc} compared to measured $L_3=4.3(1.8)\times 10^{-29}$ cm$^6$/s \cite{Burt:1997zz}}.
This scenario suggests that a lattice experiment with ${}^{87}\mathrm{Rb}$ could greatly decrease the uncertainty in the ${}^{87}\mathrm{Rb}$ three-body recombination loss coefficient and provide a better test of the theory for that system.

Finally, we study the occupation-number dependence of the lifetimes.
Fig.~\ref{fig:fig4}b shows a $\tau_{\mathrm{g}\cdots}^{-1} = \tau_\mathrm{ggg}^{-1} {n \choose 3}$ scaling of the $n$-body ground state lifetime for $n\geq3$, suggesting that three-body loss remains the dominant mechanism.
The lifetimes of the $n$-atom excited states, along with their expected scalings from counting the number of three-body loss channels, are shown in Fig~\ref{fig:fig4}c (see Methods).
These relatively long lifetimes are promising for future experiments involving coupled wells with large occupation numbers.

In conclusion, we have demonstrated two manifestations of multi-body interactions arising from pairwise interactions in few-body systems of fermions. 
Our spectroscopic technique, along with spatially resolved readout, enables efficient isolation of few-body systems, which prove to be ideal for observing multi-body effects.
It also provides a simple way to create the highly entangled and long-lived states $\ket{\mathrm{eg}\cdots^\pm}$, offering a useful resource for quantum information processing \cite{Zang:2015kb}.
The few-body systems enable precise measurements of the $a_{\mathrm{eg}^{\pm}}$ scattering lengths and the three-body loss rates that agree with the universal van der Waals model.
The collisional parameters, in the case of ${}^{87}\mathrm{Sr}$, are found to be particularly suitable for studies of two-orbital SU($N$) magnetism which should arise in the presence of weak tunneling.
These interactions have been predicted to create long-sought states of matter, including valence bond solids and chiral spin liquids \cite{Cazalilla:2014kq}.

\bibliography{bib_interaction}

\noindent
\\
\textbf{Acknowledgements} We acknowledge technical contributions from W. Milner, E. Oelker, J. Robinson, L. Sonderhouse, W. Zhang, and useful discussions with T. Bothwell, S. Bromley, C. Kennedy, D. Kedar, S. Kolkowitz, M. D. Lukin, A. Safavi-Naini, C. Sanner. This work is supported by NIST, DARPA, W911NF-16-1-0576 through ARO, AFOSR-MURI, AFOSR, NSF-1734006 and NASA. A.G. is supported by a postdoctoral fellowship from the Japan Society for the Promotion of Science and G.E.M. is supported by a postdoctoral fellowship from the National Research Council. J.P.D. acknowledges support from NSF Grant PHY-1607204.\linebreak
\linebreak
\noindent
\textbf{Author Contributions} A.G., R.B.H, G.E.M., S.L.C. and J.Y. contributed to the experiments. M.A.P., P.S.J., J.P.D and A.M.R. contributed to the development of the theoretical model. 
All authors discussed the results, contributed to the data analysis and worked together on the manuscript.\linebreak
\linebreak
\noindent
\textbf{Author Information}  The authors declare no competing financial interests. 
Readers are welcome to comment on the online version of the paper.
Correspondence and requests for materials should be addressed to A.G. (Akihisa.Goban@jila.colorado.edu) and J.Y. (Ye@jila.colorado.edu)
\newpage

\appendix
\section{Methods}
\noindent
\textbf{State preparation:}
At the end of a 10-s evaporation, a 10-spin-component Fermi gas is loaded into a cubic state-independent optical lattice in a two-stage ramp.
The first 300 ms ramp to $\sim 5$ $E_\mathrm{rec}$ is used consistently for all the measurements shown in the manuscript.
To prepare for $n=4,5$ occupied sites, the second ramp to the final lattice depth is sped up from 200 ms to 50 ms in order to minimize three-body loss during the loading process.
$n$-occupied sites are randomly filled with $n$ different nuclear-spin components among $\binom{10}{n}$ nuclear-spin configurations.
The initial entropy per particle in the lattice is estimated to be $s/k_B=1.8$ from the measured $T/T_F$ in the dipole trap before and after lattice loading. 
In the atomic limit where tunneling is negligible, the maximum spin entropy for $n=1$ sites is $s_{\rm spin}/k_{B}=\ln(10)=2.3$ for 10 spin states. This leads to a lowered temperature in the lattice when entropy is transferred from the motional to the spin degree of freedom~\cite{Taie:2012jy, Hofrichter:2016iq}.

To minimize a systematic shift due to the inhomogenity of trap depth across the cloud, we prepare a 10 $\mu$m $\times$ 10 $\mu$m $\times$ 2 $\mu$m sample of desired $n$-occupied sites by optimizing the final evaporation point.
For each image, we measure the spectroscopic response only in the 4 $\mu$m $\times$ 4 $\mu$m region centered at the lattice.
As the on-site frequency shifts for large occupations increase, the line shapes become asymmetric due to the residual inhomogeneity of the trap depth.
To determine the peak frequencies, we fit each spectrum by asymmetric Lorentzian as shown in Fig. \ref{fig:fig2}a.
The trap depth in the central region of the lattice is calibrated by motional sideband spectroscopy of a $n=1$ sample with the same procedure.
\linebreak
\linebreak
\noindent
\textbf{Low-energy effective multi-body theory:}
To obtain a concise description of the low-energy physics by integrating out higher motional states, it is convenient to carry out renormalized perturbation theory as previously studied in Bose gases trapped in a harmonic trap \cite{Johnson:2012cl}.
In our case, the existence of four scattering lengths and the fermionic character of the atoms increase the complexity of low-energy collisional physics.
Here, we briefly describe the derivation of effective two- and three-body interactions from bare two-body interactions and refer the detailed discussion to the Ref. \cite{Perlin:2017va}.

When atoms are localized in a single lattice site, a bare two-body interaction Hamiltonian consists of ground-state, excited-state, direct and exchange interactions as \cite{Gorshkov:2010hw}
\begin{align}
H_{\mathrm{int}}
&= \frac{4\pi\hbar^2}{m_a} \sum_{x\in\{\mathrm{g}, \mathrm{e}\}}\sum_{m<m'} a_{xx} \int d^3 {\bf r}\rho_{x, m}\rho_{x, m'}\nonumber\\
&+\frac{2\pi\hbar^2}{m_a} \left(a_{\mathrm{eg}^+}+a_{\mathrm{eg}^-}\right) \sum_{m, m'} \int d^3 {\bf r}\rho_{\mathrm{e}, m}\rho_{\mathrm{g}, m'}\\
&+\frac{2\pi\hbar^2}{m_a} \left(a_{\mathrm{eg}^+}-a_{\mathrm{eg}^-}\right)  \sum_{m, m'} \int d^3 {\bf r}\psi_{\mathrm{g}, m}^\dagger\psi_{\mathrm{e}, m'}^\dagger\psi_{\mathrm{g}, m'}\psi_{\mathrm{e}, m},\nonumber
\label{eq:bare-hamiltonian}
\end{align}
with a fermionic field operator $\psi_{x, m}$ for atoms in the orbital $x$ and nuclear spin $m\in\{-I, -I+1, \ldots, I\}$ and density operator $\rho_{x, m}=\psi_{x, m}^\dagger \psi_{x, m}$.
The field operator $\psi_{x, m}$ is expanded in the Wannier basis $\phi_{\alpha}({\bf r})$ for a single lattice site as $\psi_{x, m}=\sum_{\alpha} \phi_{\alpha}({\bf r})c_{\alpha, x, m}$ where $c_{\alpha, x, m}$ annihilate an atom with a motional state $\alpha$, orbital $x$, and spin $m$.

To integrate out higher-motional states, we introduce a projection operator onto the single-particle motional ground states $\mathcal{P}_0$ and an operator $\mathcal{I}\equiv \sum_{\ket\alpha \neq \ket 0}\frac{\op\alpha}{E_\alpha}$ which sums over projections onto higher-motional states $\ket\alpha$ with corresponding motional energy $E_\alpha$ where we fix the lowest-motional state energy $E_0=0$.
Then, the $i$-th order terms $H^{(i)}_{\mathrm{int}}$ in perturbative expansion up to the third order are written by \cite{Perlin:2017va}
\begin{eqnarray}
    H_{\mathrm{int}}^{(1)}
  &=& \mathcal{P}_0 H_{\mathrm{int}} \mathcal{P}_0\label{eqn:first_perturbation}\\
  H_{\mathrm{int}}^{(2)}
  &=& -\mathcal{P}_0 H_{\mathrm{int}}
  \mathcal{I} H_{\mathrm{int}} \mathcal{P}_0\label{eqn:second_perturbation}\\
  H_{\mathrm{int}}^{(3)}
 &=& \mathcal{P}_0 H_{\mathrm{int}} \mathcal{I}
  H_{\mathrm{int}} \mathcal{I} H_{\mathrm{int}} \mathcal{P}_0 \nonumber\\
  && - \frac12 \left[\mathcal{P}_0 H_{\mathrm{int}} \mathcal{P}_0,
    \mathcal{P}_0 H_{\mathrm{int}} \mathcal{I}^2
    H_{\mathrm{int}} \mathcal{P}_0\right]_+,
\label{eqn:third_perturbation}
\end{eqnarray}
where $[X,Y]_+ \equiv XY+YX$.
We note that to compute $\mathcal{I}$ we solve for the exact single particle motional states and eigenenergies in the lattice.
When we consider the case where each atom is in a different spin state, the equations (\ref{eqn:first_perturbation}-\ref{eqn:third_perturbation}) leads to effective two-body Hamiltonian $H$ in equation (\ref{eqn:two-body-hamiltonian}) with renormalization of the bare scattering lengths by \textit{in-trap} scattering lengths, where we henceforth drop the subscript of motional states, $c_{\alpha=0, x, m}\rightarrow c_{x, m}$, without consequence.
Considering at most one orbital excitation for $n$-occupied sites and unit occupation of nuclear spin states, the second- and third-order terms,  $H_{\mathrm{int}}^{(2)}$ and $H_{\mathrm{int}}^{(3)}$, give rise to effective multi-body Hamiltonian $H'$ in equation (\ref{eqn:multi-body-hamiltonian}).  
Then, the total interaction energy is given by a sum of effective two- and multi-body interactions of $H+H'$.
Note that the third-order terms in equation (\ref{eqn:third_perturbation}) also include effective four-body interactions, but their contribution is so small that they can be ignored. 
This is the reason why we omit four-body terms in $H'$ and consider only three-body terms for the calculated shifts shown in Fig. \ref{fig:fig3}.

The effective two-body Hamiltonian $H$ has a ground state
\begin{equation}
\ket{\mathrm{g}\cdots} = \prod_m c_{\mathrm{g}, m}^\dagger \ket{\t{vacuum}},
\end{equation}
where $m$ spans $n$ different nuclear-spin states chosen from $\{-9/2, \cdots, 9/2\}$ and two distinct eigenenergies with a single excitation.
One of them is associated to an orbital-symmetric excited state
\begin{equation}
\ket{\mathrm{eg}\cdots^+} = \frac1{\sqrt n} \sum_m c_{\mathrm{e}, m}^\dag c_{\mathrm{g}, m} \ket{\mathrm{g}\cdots},
\end{equation}
and the other one is associated to $(n-1)$-fold degenerate excited states $\ket{\mathrm{eg}\cdots^-}$ which are linear combinations of the states
\begin{equation}
\ket{\mathrm{eg}\cdots^-_j} = \frac1{\sqrt2} \p{c_{\mathrm{e}, m_1}^\dag c_{\mathrm{g}, m_1} - c_{\mathrm{e}, m_j}^\dag c_{\mathrm{g}, m_j}} \ket{\mathrm{g}\cdots},
\end{equation}
for $j\in\{2,\ldots,n\}$.
As fermionic statistics requires the total state to be antisymmetric, the states $\ket{\mathrm{g}\cdots}$ and $\ket{\mathrm{eg}\cdots^+}$ are symmetric in their orbital degrees of freedom, and therefore form an antisymmetric SU($N$) singlet in their nuclear spin degrees of freedom. 
The states $\ket{\mathrm{eg}\cdots^-}$ for $n\geq3$ are not separable between the orbital and nuclear spin degrees of freedom since each degree of freedom has a mixed symmetry.
The corresponding energy eigenvalues of $H$ are given by
\begin{equation}
\begin{aligned}
E_{\mathrm{g}\cdots}^{(2)}=&\quad{n \choose 2} U_\mathrm{gg}^{(2)}\\
E_{\mathrm{eg}\cdots^+}^{(2)}=&\quad{n-1 \choose 2} U_\mathrm{gg}^{(2)} + \frac{n-1}{2} (V^{(2)}+V_\mathrm{ex}^{(2)})\\
E_{\mathrm{eg}\cdots^-}^{(2)}=&\quad{n-1 \choose 2} U_\mathrm{gg}^{(2)} + \frac{n-1}{2}V^{(2)} - \frac{1}{2}V_\mathrm{ex}^{(2)}.
\end{aligned}
\end{equation}

Due to the SU($N$) symmetry of the two-body interactions, the effective three-body Hamiltonians $H'$ preserve eigenstates, modifying only the spectra of $n$-occupied sites as
\begin{eqnarray}
E_{\mathrm{g}\cdots}^{(3)}&=&{n\choose 3} U_\mathrm{ggg}^{(3)}\nonumber\\
E_{\mathrm{eg}\cdots^+}^{(3)}&=&{n-1 \choose 3} U_\mathrm{ggg}^{(3)} + {n-1 \choose 2}(V^{(3)}+V_{\mathrm{ex}}^{(3)}) \label{eqn:eigene_3body}\\
E_{\mathrm{eg}\cdots^-}^{(3)}&=&{n-1 \choose 3} U_{\mathrm{ggg}}^{(3)} + {n-1 \choose 2} V^{(3)} - \frac{n-2}{2}V_{\mathrm{ex}}^{(3)},\nonumber
\label{eqn:three-eigen-energy}
\end{eqnarray}
where $U_\mathrm{ggg}^{(3)}$, $V^{(3)}$ and $V_\mathrm{ex}^{(3)}$ are effective three-body ground-state, direct, and exchange interaction energies, which arise from the second- and third-order terms in the perturbative expansion in equations (\ref{eqn:second_perturbation}, \ref{eqn:third_perturbation}), to achieve better quantitative agreement with experimental data.
The frequency shifts from effective three-body interactions, $(E^{(3)}_{\mathrm{eg}\cdots^{\pm}}-E^{(3)}_{\mathrm{g}\cdots})/h$, are non-linear in occupation number $n$ due to $n^2$-dependent terms, as observed in Fig. \ref{fig:fig3}.
For numerical evaluation of effective two- and thee-body on-site interaction energies, we include the anisotropy of trap by using measured trap depths, $\mathcal{U}_x$, $\mathcal{U}_y$, and $\mathcal{U}_z$.
We note that, for our experimental conditions, approximating an isotropic trap by taking a geometric mean trap depth, $\mathcal{U}=(\mathcal{U}_x\mathcal{U}_y\mathcal{U}_z)^{1/3}$, does not produce a major difference.
A conservative estimate of the dominant correction to interaction energies from tunneling and inter-site interactions can be performed by treating these effects perturbatively and assuming no energetic penalty for nearest-neighbor hopping.  
These corrections in our theoretical calculations are estimated to be $\lesssim 1$\%.  
Extended Data Fig. \ref{fig:exfig1} shows good agreement between measured data and calculated shifts including three-body interactions at the mean trap depths from $30$ to $80$ $E_{\mathrm{rec}}$.
\setcounter{figure}{0}    
\begin{figure}[t!]
\renewcommand\figurename{Extended Data Fig.}
\centering
\includegraphics[width=1\columnwidth]{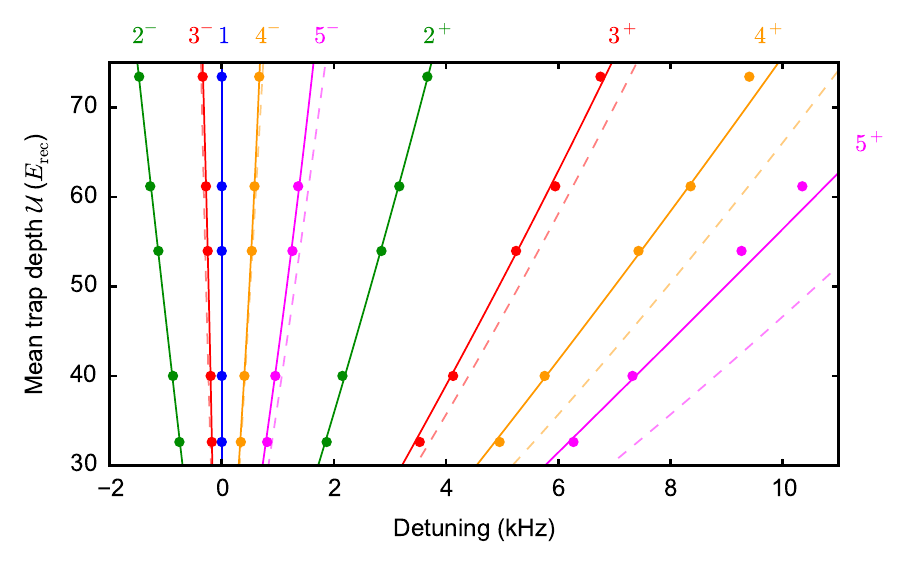}
\caption{
    \textbf{Trap-depth dependence of multi-body clock shifts.} 
   Clock shifts of $\ket{\mathrm{eg}\cdots^{\pm}}$ from $n=2,3,4, \mathrm{and}~5$, as a function of the mean trap depth.
   The labels $n^{\pm}$ denote excitation of $\ket{\mathrm{eg}\cdots^\pm}$ for $n$-occupied sites.
    The calculated shifts from the effective two-body and the three-body theories are shown in dashed and solid lines, respectively, where the calculated shifts are interpolated for a guide to the eye.
These calculated shifts from the three-body theory show good agreement with the experimental data in the wide range of the mean trap depths.
The uncertainties of the experimental data are smaller than the size of the data points.
}
\label{fig:exfig1}
\end{figure}
\linebreak
\linebreak
\noindent
\textbf{Occupation number dependence of three-body lifetime:}
Assuming that three-body loss is the dominant loss mechanism for $n\geq3$ atoms in a lattice site, we consider scalings of lifetimes for $\ket{\mathrm{g}\cdots}$ and $\ket{\mathrm{eg}\cdots^{\pm}}$ states.
The three-body loss can be taken into account by adding imaginary parts of the eigenenergies through $E^{(3)}_X\rightarrow\tilde{E}_X^{(3)}=E_X^{(3)}-i\frac{\hbar}{\tau_X}$ where $X\in\{\mathrm{ggg}, \mathrm{egg}^\pm\}$.
For $n\geq3$ atoms, by rewriting the equations (\ref{eqn:eigene_3body}) with $\tilde{E}^{(3)}_{\mathrm{ggg}}$ and $\tilde{E}^{(3)}_{\mathrm{egg}^{\pm}}$, we obtain the scaling of $\tau_{\mathrm{g}\cdots}$ and $\tau_{\mathrm{eg}\cdots^{\pm}}$ as
\begin{eqnarray}
\tau_{\mathrm{g}\cdots}^{-1} &=&  \tau_{\mathrm{ggg}}^{-1}{n\choose 3}\nonumber\\
\tau_{\mathrm{eg}\cdots^+}^{-1} & =&  \tau_{\mathrm{egg}^+}^{-1} {n-1 \choose 2} + \tau_{\mathrm{ggg}}^{-1} {n-1 \choose 3}\label{eqn:eggp}\\
\tau_{\mathrm{eg}\cdots^-}^{-1} & =& \tau_{\mathrm{egg}^-}^{-1} \frac{2n}{3n-3}{n-1 \choose 2} + \tau_{\mathrm{egg}^+}^{-1} \frac{n-3}{3n-3}{n-1 \choose 2} \nonumber\\
&&  + \tau_{\mathrm{ggg}}^{-1} {n-1 \choose 3}.\nonumber
\label{eqn:eggm}
\end{eqnarray}
\linebreak
\linebreak
\noindent
\textbf{Extracting free-space scattering lengths:}
To extract a free-space scattering length $a_X$ with $X\in\{\mathrm{eg}^\pm\}$, we need to consider the contributions from off-resonant excitations to the higher motional bands and a finite-range correction to the zero-range potential.
Here, we apply the pertubative expansions, previously studied for  bosons in a harmonic trap in Ref. \cite{Johnson:2012cl}, to fermions with different nuclear-spin states in the lattice.
By approximating our slightly anisotropic trap by an isotropic trap with a geometric mean trap frequency $\omega$, the interaction energy of doubly occupied sites in the lattice is written by the perturbative expansions in $a_X/l_0(\omega)$ with a harmonic length $l_0(\omega)=\sqrt{\frac{\hbar}{m_a\omega}}$ as
\begin{equation}
\begin{aligned}
E_X=&\sum_{i\in\{1, 2, 3\}}\tilde{c}_2^{(i)}\left(\frac{a_X}{l_0(\omega)}\right)^i
+\tilde{d}_2^{(1,2)}\left(\frac{r_{\mathrm{eff}, X}}{l_0(\omega)}\right)\left(\frac{a_X}{l_0(\omega)}\right)^2,
\label{eqn:perturbative-expansion}
\end{aligned}
\end{equation}
where $\tilde{c}_2^{(i)}$ determines the $i$-th order two-body correction due to lattice confinement and $\tilde{d}_2^{(1,2)}$ determines correction from the effective ranges of $r_{\mathrm{eff}, X}$.
We calculate $\tilde{c}_2^{(1)}$ and $\tilde{d}_2^{(1,2)}$ by using the wave function of the ground motional band of the lattice, while evaluation of $\tilde{c}_2^{(2)}$ and $\tilde{c}_2^{(3)}$ require numerical calculations which do not converge fast with increasing number of motional bands.
Therefore, to estimate $\tilde{c}_2^{(i\geq 2)}$, we rescale the ones for the harmonic trap to account for the anharmonicity of the lattice since the two-body coefficients, $c_2^{(i)}$, in the harmonic trap are analytically calculated in Ref. \cite{Johnson:2012cl}.
Explicitly, we approximate $\tilde{c}_2^{(i)}=\eta^i\cdot c_2^{(i)}$ for $i\geq 2$ with $\eta=\tilde{c}_2^{(1)}/c_2^{(1)}$, as $\tilde{c}_2^{(i)}$ contains the product of $i$ spatial overlap integrals.
The effective ranges $r_{\mathrm{eff}, X}$ are analytically calculated by considering a long-range van der Waals potential with the computed values of $C_{6, X}$ \cite{Flambaum:1999hr,Zhang:2014elb}.
By using the measured frequency shifts $(E_{\mathrm{eg}^{\pm}}-E_{\mathrm{gg}})/h$ at the mean trap depths ranging from $30$ to $80$ $E_{\mathrm{rec}}$ as shown in Extended Data Fig. \ref{fig:exfig1}, combined with the previous measurement of $a_\mathrm{gg}=96.2(0.1)$ \cite{MartinezdeEscobar:2008eu, Stein:2010dg}, we extract the free-space scattering lengths as $a_{\mathrm{eg}^+} = 160.0(0.5)_\mathrm{stat}(2.3)_\mathrm{sys}~a_0$ and $a_{\mathrm{eg}^-} = 69.1(0.2)_\mathrm{stat}(0.9)_\mathrm{sys}~a_0$, summarized in Table \ref{tab:experiments}.
Due to the approximate nature of this treatment, we conservatively consider the second- and third-order corrections parametrized by $\tilde{c}_2^{(i\geq2)}$ in equation (\ref{eqn:perturbative-expansion}) as systematic uncertainties.
\linebreak
\linebreak
\noindent
\textbf{Calculations of three-body lifetimes based on a universal van der Waals model:}
Our numerical three-body calculation uses a universal van der Waals model in an adiabatic hyperspherical basis 
\cite{Wang:2012bc, Wang:2014hx,Wolf:2017fc}. 
Using the computed $C_6$ \cite{Zhang:2014elb} and $C_8$ \cite{Porsev:2006ij, Porsev:2014jr}  coefficients of the ${}^1\mathrm{S}_0+{}^1\mathrm{S}_0$ and ${}^1\mathrm{S}_0+{}^3\mathrm{P}_0$ scattering channels \cite{Zhang:2014elb} in our two-body potential model, we determined the lifetime of three-body states by calculating the complex eigenvalue of three atoms in an isotropic harmonic trap whose trapping frequency reproduces the value of the zero-point energy of the experimental trap configuration for a given value of the mean trap depth (A complex eigenvalue is obtained by leaving the adiabatic channels associated to the diatomic states open, thus allowing the state to decay).
The imaginary part of the eigenvalue yields the lifetime. In our calculations we treat the $\ket{\mathrm{ggg}}$ and 
$\ket{\mathrm{egg}^+}$ states as states of a system with three- and two-identical bosons, respectively, owing to the antisymmetric character of the spin components of such states. 
Due to the mixed character of the spins in the $\ket{\mathrm{egg}^{-}}$ states, we treat such states as states of a system with three dissimilar atoms, whose interactions are determined from those of the $\ket{\mathrm{gg}}$, $\ket{\mathrm{eg}^+}$, and $\ket{\mathrm{eg}^-}$ states.
In our studies, the lifetimes were obtained for potentials supporting an increasing number of diatomic molecular states and are, similarly to the finding in Ref.~\cite{Wolf:2017fc}, shown to quickly converge.
Our results with a model potential containing three $s$-wave bound states is converged within $<$10\%.
These calculations were performed keeping the values of the two-body scattering lengths fixed to their known values (see Table \ref{tab:experiments}).
Our analysis indicates that the substantially higher values for the three-body loss coefficients $\beta_{\mathrm{egg}^{\pm}}$ in comparison to $\beta_{\mathrm{ggg}}$ can be attributed to the increase on number of possible decaying channels available to the $\ket{\mathrm{egg}^{\pm}}$ states.
In fact, while $\ket{\mathrm{ggg}}$ states can only decay into $\ket{\mathrm{gg}}$ molecular states, $\ket{\mathrm{egg}^+}$ states
can decay to $\ket{\mathrm{gg}}$ and $\ket{\mathrm{eg}^+}$ molecules, and $\ket{\mathrm{egg}^-}$ states can decay to all 
possible molecular states, $\ket{\mathrm{gg}}$, $\ket{\mathrm{eg}^+}$ and $\ket{\mathrm{eg}^-}$.
We note that our model for $\ket{\mathrm{egg}^+}$ states only allows for decay into $\ket{\mathrm{eg}}$ and $\ket{\mathrm{eg}^+}$ states, while in 
a spin dependent model \cite{Wang:2014hx} should allow for additional decay to $\ket{\mathrm{eg}^-}$ states. As a consequence, our model for $\ket{\mathrm{egg}^+}$ states overestimate their lifetimes.
\linebreak
\linebreak
\noindent
\textbf{Data availability:} The data supporting the findings of this study are available within the paper and its Extended Data.

\end{document}